\documentclass[aps,prc,twocolumn,showpacs,floatfix,superscriptaddress,amsmath,amssymb,nofootinbib,longbibliography]{revtex4-1}

\usepackage{hyperref}
\usepackage{color}
\usepackage{amsmath}

\newcommand{\be}{\begin{equation}}
\newcommand{\ee}{  \end{equation}}
\newcommand{\ba}{\begin{eqnarray}}
\newcommand{\ea}{  \end{eqnarray}}
\newcommand{\bas}{\begin{eqnarray*}}
\newcommand{\eas}{  \end{eqnarray*}}

\begin{document}

\title{Transport Equations for Driven Many-body Quantum Systems}

\author{H. A. \surname{Weidenm{\"u}ller}}
\email{haw@mpi-hd.mpg.de}

\affiliation{Max-Planck-Institut f{\"u}r Kernphysik, D-69029
  Heidelberg, Germany}

\begin{abstract}Transport equations for autonomous driven fermionic
  quantum systems are derived with the help of statistical
  assumptions, and of the Markov approximation. The statistical
  assumptions hold if the system consists of subsystems within which
  equilibration is sufficiently fast. The Markov approximation holds
  if the level density in each subsystem is sufficiently smooth in
  energy. The transport equation describes both, relaxation of
  occupation probability among subsytems at equal energy that leads to
  thermalization, and the transport of the system to higher energy
  caused by the driving force. The laser-nucleus interaction serves as
  an example for the applicability and flexibility of the approach.
\end{abstract}


\maketitle

\section{Introduction}
\label{int}

We consider autonomous strongly interacting fermion systems under the
influence of an external driving force. Examples are atomic nuclei hit
by a strong laser pulse carrying MeV photons, and atoms in a trap
where the confining potential oscillates with time. In both cases the
motion of the constituents (nucleons or fermionic atoms) is primarily
determined by a stationary potential (the mean field of the nuclear
shell model or the stationary confining potential of the trap) which
binds the particles. The external time-dependent force (the laser
pulse or the oscillating potential) drives either system towards
higher excitation energies. The interaction between the constituents
(nucleons or atoms) drives the system towards statistical equilibrium.

Normally, a physical system equilibrates because it is coupled to an
external reservoir (a heat bath). If driven by an external force, such
a system is described in terms of a transport equation (a standard
tool of nonequilibrium statistical quantum mechanics). The
justification for that equation, based upon the presence of a heat
bath, fails for autonomous systems. Transport equations also hold for
autonomous systems. A prime example is the quantum Boltzmann equation
that describes a quantum gas of interacting particles. Here
equilibration is caused by the collision term describing particle
interaction. The standard derivation of the quantum Boltzmann equation
uses methods of quantum field theory~\cite{Sno12}. The systems we
consider differ from the Boltzmann gas by the presence of a strong
stationary potential. Transport equations have been used also in that
context, for instance for the laser-nucleus interaction~\cite{Pal14,
  Pal15, Kob20}, albeit without firm theoretical justification. In the
present paper we supply that justification. We show that statistical
assumptions on coupling matrix elements between constituent parts of
the system take the role of the heat bath. We derive the transport
equations and, thereby, display the conditions of validity of these
equations for autonomous systems.

In Section~\ref{exam} we give two examples where transport equations
have been or may be useful. In Section~\ref{mar} we introduce and
justify our statistical assumptions, and we define the conditions that
allow us to use the Markov approximation in deriving the transport
equation. In Section~\ref{rate} we derive and discuss the transport
equation for internal relaxation. In Section~\ref{time} we
generalize the approach to the description of driven
systems. Section~\ref{nda} is devoted specifically to nuclear dipole
absorption. Section~\ref{sum} contains a brief summary. Technical
details are relegated to an appendix.

\section{Two Examples}
\label{exam}

The following two examples serve to set the stage and to give physical
substance to the theoretical developments in the following Sections.

\subsection{Laser-Nucleus Interaction}
\label{las}

A laser pulse carrying photons with energy $\hbar \omega_0 \approx
10$ MeV that hits a target nucleus predominantly causes dipole
excitation because the product of wave number $k$ and nuclear radius
$R$ obeys $k R \ll 1$. When the target nucleus is in the ground state,
dipole absorption excites the nuclear Giant Dipole Resonance (GDR).
Pictorically speaking, the GDR is an oscillation of the center of mass
of the protons against that of the neutrons. The Brink-Axel
hypothesis~\cite{Bri55, Axe62} postulates that this process is
universal: Dipole excitation of any initial nuclear state $| i
\rangle$ (ground or excited) with energy $E_i$, predominantly
populates the nuclear Giant Dipole Resonance (GDR) built upon that
state. The GDR is a mode $| d(i) \rangle$ of excitation given by the
normalized product of the dipole operator and the initial state $| i
\rangle$, with mean excitation energy $E_d - E_i$ where $E_d = \langle
d(i) | H_N | d(i) \rangle$. The GDR is not an eigenstate of the
nuclear Hamiltonian $H_N$. Rather, the GDR is distributed over the
eigenstates of $H_N$ with an approximately Lorentzian distribution
with width $\Gamma^\downarrow \approx 5$ MeV. In a time-dependent
picture, the ``spreading width'' $\Gamma^\downarrow$ describes the
damping of the GDR due to its rapid mixing with the eigenstates of
$H_N$. We refer to such mixing as to (nuclear) equilibration.

A sufficiently intense laser pulse may cause multiple dipole
excitation, each dipole absorption process populating the GDR built
upon the excited state(s) reached in the previous absorption
process. Every such photon absorption process is followed by partial
or complete equilibration, depending on the ratio of the rate for
dipole absorption (derived in Ref.~\cite{Pal20}) and the rate
$\Gamma^\downarrow / \hbar$ for equilibration. For the theoretical
description of the sequence of alternating absorption and
equilibration processes, the Schr{\"o}dinger equation is useless
(nuclear level densities attain enormous values already several $10$
MeV above the ground state, so that a numerical treatment is out of
the question), and transport equations are called for. These describe
the evolution in time of mean occupation probabilities of (classes of)
nuclear states, see Refs.~\cite{Pal14, Pal15, Kob20}.

Intense pulses carrying photons with energy in the MeV range are under
development both at the Nuclear Pillar of the Extreme Light
Infrastructure~\cite{ELI} and at the Gamma Factory of the Large Hadron
Collider at CERN~\cite{Pla19}. The theoretical results in
Refs.~\cite{Pal14, Pal15, Kob20} should soon encounter experimental
tests.

\subsection{Fermionic Atoms in a Trap}

We consider $N \gg 1$ fermionic atoms in a two-dimensional harmonic
oscillator. That situation can be realized experimentally by capturing
atoms with half-integer spin in a
trap~\cite{Bay20}. Two-dimensionality of the trap is effectively
achieved by making the level spacing in the third spacial direction
sufficiently large. The confining potential oscillates
harmonically. That gives rise to a time-dependent dipole-type driving
force leading to multiple dipole excitation, similar to the nuclear
case. The atom-atom interaction is controlled via a Feshbach resonance
and may be either attractive or repulsive~\cite{Chi10}. Provided it is
sufficiently strong and the excitation energy is sufficiently large,
we expect the atom-atom interaction to cause equilibration,
independently of its sign. Again, we deal with the interplay of a
dipole interaction driving the system towards higher energies, and of
the tendency of the system to equilibrate. The arguments in favor of a
theoretical description in terms of transport equations are the same
as in the nuclear case. Even for a small number of particles in the
trap, the level density grows with excitation energy so strongly that
there is no alternative to that approach. To the best of our
knowledge, equilibration processes as investigated theoretically in
this paper have not yet been studied experimentally for atoms in a
trap. The system we study is close to equilibrium and, thus, differs
from the ones studied in Refs.~\cite{Ber08, Pru20, Zac20}.

\section{Statistical Assumptions. Markov Approximation}
\label{mar}

In this Section we give the general argument used to derive
time-dependent transport equations for a time-reversal-invariant
fermionic many-body Hamiltonian $H$ of the form
\ba
\label{r0}
H = H_0 + V + W(t) \ .
\ea
Here $H_0$ is a single-particle Hamiltonian containing the kinetic
energy and a stationary central potential, $V$ is a two-body
interaction, and $W(t)$ is the time-dependent external dipole
interaction that drives the system. For nuclei, the central potential
is the shell-model potential. For atoms in a trap, it is the external
potential defining the trap plus possibly a mean-field contribution
due to the atom-atom interaction. In nuclei, the two-body interaction
$V$ is the residual interaction of the shell model. For atoms in a
trap, it is the atom-atom interaction after the mean-field
contribution has been removed. We assume that for both, nuclei and
atoms in a trap, $V$ is sufficiently weak to leave intact the basic
shell structure defined by $H_0$. For both nuclei and atoms, the
dipole interaction $W(t)$ is the sum of single-particle operators. We
consider a system of $N \gg 1$ Fermions. In nuclei these are protons
and neutrons. For atoms in a trap we assume that these have
half-integer spin.

Transport equations are derived by averaging occupation probabilities
of classes of states over the statistical distribution of matrix
elements. We follow Ref.~\cite{Aga75} where analogous arguments have
been used for the statistical treatment of the time-independent
compound-nucleus scattering cross section.

\subsection{Statistical Assumptions}
\label{stat}

The eigenstates of $H_0$ are Slater determinants. For simplicity we
assume that the ground-state Slater determinant $S_0$ corresponds to a
closed shell. Then $S_0$ has spin zero, positive parity, and defines
the Fermi energy. We define classes of Slater determinants of excited
states $(\alpha, \mu)$. The class label $\alpha$ stands for spin,
parity, particle-hole number (defined with respect to the Fermi
energy), and excitation energy while $\mu$ is a running index for the
states within the class.

Our basic statistical assumption is formulated in terms of the
time-independent part $H_0 + V$ of $H$. We choose a representation
where $H_0 + V$ is diagonalized within each class, with eigenvalues
$E_{\alpha \mu}$ and eigenfunctions $| \alpha \mu \rangle$. If the
number of Slater determinants in each class is sufficiently large, and
if the mixing of determinants in that class due to $V$ is sufficiently
thorough, it is justified to assume that within that class, $H_0 + V$
acts like a random Hamiltonian, a member of the Gaussian Orthogonal
Ensemble (GOE) of random matrices. Then for each class $\alpha$ the
projections of the eigenfunctions $| \alpha \mu \rangle$ onto some
fixed state are Gaussian-distributed random variables, and the
eigenvalues $E_{\alpha \mu}$ obey Wigner-Dyson statistics. That is our
basic statistical assumption. It plays the role of the heat bath for
open systems.

Our assumption neglects the existence of specific, nonstatistical
modes of excitation that occur both in nuclei and for atoms in a
trap. In nuclei, these are, for instance, collective modes associated
with quadrupole deformations of the ground state~\cite{BM}. In traps,
the Higgs mode~\cite{Bay20} is an example. However, such modes are
expected to be washed out at the higher excitation energies relevant
for the present paper and are, therefore, disregarded.

The most thorough numerical test of our statistical assumption that we
are aware of was performed in Ref.~\cite{Zel}. Spectra and
eigenfunctions of nuclei in the middle of the $s d$-shell were
calculated in the framework of the nuclear shell model with a residual
two-body interaction. Typical dimensions of the Hamiltonian matrices
were of the order of $10^3$. In the centres of the spectra so
obtained, level spacing distribution and eigenfunction statistics
agreed well with GOE predictions. We assume that the same statements
hold for the eigenvalues $E_{\alpha \mu}$ and eigenfunctions $| \alpha
\mu \rangle$ within each class $\alpha$.

In the context of our time-dependent transport equation, we
distinguish equilibration and relaxation. Our statistical assumption
implies that within each class, equilibration is instantaneous or,
physically more accurately, that the equilibration time within each
class is short in comparison with the other characteristic time scales
of the system. These are the time scale for on-shell relaxation of
occupation probabilities in different classes, and the time scale for
dipole excitation. Both time scales enter the transport equation(s).
Obviously, in every application of the formalism the classes $\alpha$
must be chosen judiciously so as to approximately satisfy that
assumption. In nuclear physics empirical evidence supports our scheme.
Precompound reactions initiated by light projectiles (protons or alpha
particles) are described successfully as a sequence of nucleon-nucleon
collisions, each collision populating the class of next-higher
particle-hole number~\cite{Bla75}.

States in different classes $\alpha \neq \beta$ are connected by the
nondiagonal elements $\langle \beta \nu | V | \alpha \mu \rangle$ of
$H$. Because of time-reversal invariance these can be chosen real and
symmetric. Obviously, the matrix elements of $V$ vanish unless the
classes $\alpha$ and $\beta$ carry the same conserved quantum
numbers. Since $V$ is a two-body interaction, the elements $\langle
\beta \nu | V | \alpha \mu \rangle$ also vanish if $\alpha$ and
$\beta$ differ in particle-hole number by more than one unit. Because
of our statistical assumption, the non-vanishing non-diagonal matrix
elements $\langle \beta \nu | V | \alpha \mu \rangle$ are
zero-centered Gaussian-distributed random variables with second moments
given by
\ba
&& \bigg\langle \langle \beta \nu | V | \alpha \mu \rangle \langle
\beta' \nu' | V | \alpha' \mu' \rangle \bigg\rangle = \bigg(
\delta_{\alpha \alpha'} \delta_{\mu \mu'} \delta_{\beta \beta'} \delta_{\nu \nu'}
\nonumber \\
&& \qquad \qquad \qquad + \delta_{\alpha \beta'} \delta_{\mu \nu'}
\delta_{\beta \alpha'} \delta_{\nu \mu'} \bigg) V^2_{\beta \alpha} \ .
\label{r1}
\ea
Here and in what follows, big angular brackets denote the average over
the Gaussian distribution of matrix elements. The right-hand side of
Eq.~(\ref{r1}) defines the second moment $V^2_{\alpha \beta} =
V^2_{\alpha \beta}$. The second moment is equal to the mean square
matrix element of $V$ connecting states in classes $\alpha$ and
$\beta$ and, therefore, measures the strength of the coupling between
the two classes. The matrix elements $\langle \alpha \mu | W(t) |
\beta \nu \rangle$ of the dipole interaction vanish unless the classes
$\alpha$ and $\beta$ differ in parity. These elements, too, are
zero-centered Gaussian-distributed random variables. They are
addressed in Section~\ref{time} below.

\subsection{Markov Approximation}
\label{mark}

We introduce the Markov approximation for the time-independent part
$H_0 + V$ of the Hamiltonian $H$. To that end we calculate the time
dependence of the mean total occupation probability $P_\alpha(t)$ of
the states in class $\alpha$. The mean value is taken over the
Gaussian distribution of the coupling matrix elements~(\ref{r1}).
Keeping only the terms of leading order (defined below) in that
average defines the Markov approximation. The procedure applies
analogously to the time-dependent part $W(t)$ of the full
Hamiltonian~(\ref{r0}).

The solution $\Psi(t)$ of the time-dependent Schr{\"o}dinger equation
\ba
\label{ref1}
i \hbar \frac{\partial}{\partial t} \Psi(t) = (H_0 + V) \Psi(t)
\ea
conserves probability, $({\rm d} / {\rm d} t) |\Psi(t)|^2 = 0$,
because $H_0 + V$ is Hermitean. In view of the statistical
assumptions~(\ref{r1}), the elements of the Hamiltonian $H_0 + V$ form
a random-matrix ensemble, and $\Psi(t)$ is a random variable. For
every realization of $H_0 + V$ the solution $\Psi(t)$ conserves
probability. The same is then true of the average probability,
\ba
\frac{\rm d}{{\rm d} t} \bigg\langle | \Psi(t) |^2 \bigg\rangle = 0 \ .
\label{r2}
\ea
We use the interaction representation and expand $\Psi(t)$ in the
basis of states $| \alpha \mu \rangle$,
\ba
| \Psi(t) \rangle = \sum_{\alpha \mu} c_{\alpha \mu}(t) \exp \{ - i
E_{ \alpha \mu} t / \hbar \} | \alpha \mu \rangle \ .
\label{r3}
\ea
We recall that the states $|\alpha \mu \rangle$ are eigenstates of the
projection of $(H_0 + V)$ onto the space spanned by the states in
class $\alpha$. The time dependence of $c_\alpha(t)$ is due to
elements of $V$ in Eq.~(\ref{ref1}) that connect states in classes
$\alpha \neq \beta$. The average total occupation probability
$P_\alpha(t)$ of states in group $\alpha$ at time $t$ is defined as
\ba
P_\alpha(t) = \sum_\mu \bigg\langle | c_{\alpha \mu}(t) |^2 \bigg\rangle \ .
\label{r4}
\ea
From Eq.~(\ref{r3}) we have $\langle \Psi(t) | \Psi(t) \rangle =
\sum_{\alpha \mu} | c_{\alpha \mu}(t) |^2$, and Eq.~(\ref{r2})
combined with the definition~(\ref{r4}) yields
\ba
\frac{\rm d}{{\rm d} t} \sum_\alpha P_\alpha(t) = 0 \ .
\label{r5}
\ea
Total probability (i.e., the sum of the average total occupation
probabilities of the states $(\alpha, \mu)$) is conserved.

In the interaction representation, the time-dependent Schr{\"o}dinger
equation reads
\ba
\label{r6}
i \hbar \dot{c}_{\alpha \mu}(t) = \sum_{\beta \nu} \tilde{V}_{\alpha \mu;
  \beta \nu}(t) c_{\beta \nu}(t) \ .
\ea
The elements of the Hermitean matrix $\tilde{V}$ are given by
\ba \tilde{V}_{\alpha \mu; \beta \nu}(t) = \langle \alpha \mu | V | \beta \nu
\rangle \exp \{ i (E_{\alpha \mu} - E_{\beta \nu}) t / \hbar \} \ .
\label{r7}
\ea
Here $\alpha \neq \beta$. The Gaussian distribution of the matrix
elements of $V$ implies that the matrix elements $\tilde{V}_{\alpha
  \mu; \beta \nu}$ are likewise zero-centered Gaussian random
variables with second moment (see Eq.~(\ref{r1}))
\ba
&& \langle \tilde{V}_{\alpha \mu; \beta \nu}(t_1) \tilde{V}_{\alpha' \mu';
  \beta' \nu'}(t_2) \rangle = V^2_{\alpha \beta} \nonumber \\
&& \times \exp \{ i (E_{\alpha \mu} - E_{\beta \nu}) t_1 \} \exp \{ i
(E_{\alpha' \mu'} - E_{\beta' \nu'}) t_2 \} \nonumber \\
&& \times \bigg( \delta_{\alpha \alpha'} \delta_{\mu \mu'}
\delta_{\beta \beta'} \delta_{\nu \nu'} + \delta_{\alpha \beta'}
\delta_{\mu \nu'} \delta_{\beta \alpha'} \delta_{\nu \mu'} \bigg) \ .
\label{r8}
\ea
If at time $t = 0$ the nucleus is in state $| \alpha_0 \mu_0 \rangle$,
Eq.~(\ref{r6}) is supplemented by the initial condition
\ba
\label{r9}
c_{\alpha \mu}(0) = \delta_{\alpha \alpha_0} \delta_{\mu \mu_0} \ . 
\ea
Expanding the solution of Eq.~(\ref{r6}) in a formal series in powers
of $\tilde{V}$ and using Eq.~(\ref{r9}) yields
\ba
&& c_{\alpha \mu}(t) = \sum_{n = 0}^\infty {\cal T} \int_0^t \prod_{k = 1}^n
\frac{{\rm d} t_k}{i \hbar} \ \bigg( \prod_{l = 1}^n \tilde{V}(t_l)
\bigg)_{\alpha \mu; \alpha_0 \mu_0} \ .
\label{r10}
\ea
The symbol ${\cal T}$ denotes time ordering: The integration variables
obey $t \geq t_1 \geq t_2 \geq \ldots \geq t_n$. The product in big
round brackets is understood as the matrix product. For $n = 2$, for
instance, the round bracket reads $\sum_{\beta \nu} \tilde{V}_{\alpha
  \mu; \beta \nu}(t_1) \tilde{V}_{\beta \nu; \alpha_0
  \mu_0}(t_2)$. The transport equations are differential equations in
time for the mean occupation probabilities $P_\alpha$(t), subject to
the initial condition~(\ref{r9}). They are obtained by using the
expansion~(\ref{r10}) for $c_{\alpha \mu}$ and the corresponding
expansion for $c^*_{\alpha \mu}$, carrying out the average over the
Gaussian distribution of the matrix elements $\tilde{V}$, and by
taking the time derivative of the result. Because of the Gaussian
distribution, the average is obtained by summing over all ways of
averaging pairs of matrix elements of $\tilde{V}$ (``contracting''
pairs). Thus, only terms containing even powers of $\tilde{V}$
contribute to the average. The two members of a contracted pair may
both belong to $c_{\alpha \mu}$, may both belong to $c^*_{\alpha
  \mu}$, or one member may belong to $c_{\alpha \mu}$ and the other to
$c^*_{\alpha \mu}$. The number of possible contraction patterns
increases dramatically with the number $n$ of factors $\tilde{V}$ in
the series~(\ref{r10}). In praxis the average can be carried out only
if some approximation is made.

Our central assumption is that for every class $\alpha$, the average
level density $\rho_\alpha$ is very smooth. More precisely: The energy
interval over which the average level density $\rho_\alpha$ of any of
the classes $\alpha$ changes significantly, is large compared to
$\hbar$ times the inverse characteristic time $t$ for either
relaxation or for dipole absorption. That means
\ba
\label{r10a}
\frac{\hbar^n}{t^n} \frac{\partial^n \rho_\alpha}{\partial E^n} \ll
\rho_\alpha
\ea
for $n = 1, 2, \ldots$. The leading-order contraction patterns are
obtained by neglecting all energy derivatives of $\rho_\alpha$ for all
classes $\alpha$. That neglect leads to transport equations that
relate the time derivatives of the occupation probabilities
$P_\alpha(t)$ with the values of the occupation probabilities
$P_\beta(t)$ at the same time $t$ and, thus, do not carry any memory
effects. That is why we refer to that neglect as to the Markov
approximation.

\section{Transport Equation for Internal Relaxation}
\label{rate}

Using conditions~(\ref{r10a}), i.e., the Markov approximation, we now
derive the transport equation for internal relaxation, i.e., for a
system without external driving force, and we discuss the result.

\subsection{Derivation}
\label{deri}

From Eq.~(\ref{r4}) we have
\ba
\dot{P}_\alpha = \sum_\mu \bigg\langle c^*_{\alpha \mu} \dot{c}_{\alpha \mu}
\bigg\rangle + {\rm c.c.}
\label{r11}
\ea
The dots indicate time derivatives. For $\dot{c}$ we use
Eq.~(\ref{r10}) and obtain
\ba
\label{r11a}
\dot{c}_{\alpha \mu} &=& \frac{1}{i \hbar} \sum_{\beta \nu}
\tilde{V}_{\alpha \mu, \beta \nu}(t) \sum_{n = 0}^\infty {\cal T}
\int_0^t \prod_{k = 1}^n \frac{{\rm d} t_k}{i \hbar} \nonumber \\
&& \times \bigg( \prod_{l = 1}^n \tilde{V}(t_l) \bigg)_{\beta \nu;
  \alpha_0 \mu_0} \ .
\ea
For $c^*_{\alpha \mu}$ we take the complex conjugate of the
expansion~(\ref{r10}) and use $\tilde{V}^\dag = \tilde{V}$ or,
equivalently, $\tilde{V}^* = \tilde{V}^T$. The resulting expression
has the form of Eq.~(\ref{r10}) except that the sequence of factors in
the big round brackets and of the indices on these brackets is
reversed, and that $i \to - i$. Thus, the last factor on the right of
these brackets is $\tilde{V}(t_1)_{\beta \nu, \alpha \mu}$. We use
that form of $c^*_{\alpha \mu}$ in the first term on the right-hand
side of Eq.~(\ref{r11}). To calculate the average (angular brackets in
Eq.~(\ref{r11})), we start with the first factor $\tilde{V}(t)$ in the
expansion~(\ref{r11a}). According to the rules established in
Section~\ref{mark}, that factor must be contracted with some other
factor $\tilde{V}$ in the expansion of either $\dot{c}$ or of
$c^*_{\alpha \mu}$. In the appendix we show that the terms of leading
order in the sense of the inequalities~(\ref{r10a}) are obtained by
contracting that first factor $\tilde{V}(t)$ either with the factor
$\tilde{V}$ immediately to its right in the expansion of $\dot{c}$, or
with the factor $\tilde{V}$ immediately to its left in the expansion
of $c^*_{\alpha \mu}$. Using Eqs.~(\ref{r8}) we calculate both
contributions.

Contraction of the factor $\tilde{V}(t)$ with the factor immediately
to its right gives for the first term on the right-hand side of
Eq.~(\ref{r11})
\ba
\label{r11b}
&& \frac{1}{(i \hbar)^2} \sum_\beta V^2_{\alpha \beta} \sum_{\mu \nu}
\int_0^t {\rm d} t_1 \ \exp \{ i (E_{\alpha \mu} - E_{\beta \nu}) (t -
t_1) \} \nonumber \\
&& \qquad \times \bigg\langle c^*_{\alpha \mu}(t) c_{\alpha \mu}(t_1)
\bigg\rangle \ .
\ea
Here, $V^2_{\alpha \beta}$ is defined in Eq.~(\ref{r1}). We expand
$c_{\alpha \mu}(t_1)$ in a Taylor series in $t_1$. We define
$\Delta_\nu = - i (E_{\alpha \mu} - E_{\beta \nu}) / \hbar$. The
integral over the term of order $t^k_1$ is
\ba
\label{r11c}
&& \sum_\nu \int_0^t {\rm d} t_1 \ t^k_1 \exp \{ \Delta_\nu t_1) \} 
\nonumber \\
&& \qquad = \sum_\nu \frac{{\rm d}^k}{ ( {\rm d} \Delta_\nu)^k}
\int_0^t {\rm d} t_1 \ \exp \{ \Delta_\nu t_1 \} \nonumber \\
&& \qquad = \sum_\nu \frac{{\rm d}^k}{( {\rm d} \Delta_\nu)^k} \bigg[
  \frac{1} {\Delta_\nu} [ \exp \{ \Delta_\nu t \} - 1] \bigg]
\nonumber \\
&& \qquad = \sum_\nu \frac{t^k}{\Delta_\nu} \exp \{ \Delta_\nu t \}
\bigg[1 - \frac{k}{\Delta_\nu t} + \ldots \bigg] \ .
\ea
The dots indicate terms that contain factors $(\Delta_\nu t)^{- l}$
with $l \geq 2$. As shown below, criterion~(\ref{r10a}) implies that
all terms in the big straight brackets in the last line containing
factors $t^{- l} \Delta^{- l}_\nu$ with $l \geq 1$ are negigible. Thus,
the big straight brackets in the last line are replaced by unity.
Inserting the result into expression~(\ref{r11b}) we see that the
exponential factors cancel. We use
\ba
\label{r11d}
\bigg\langle \sum_\nu \frac{1}{\Delta_\nu} \bigg\rangle &=& \bigg\langle
\sum_\nu \frac{i \hbar}{E_{\alpha \mu} - E_{\beta \nu}} \bigg\rangle
\approx \pi \hbar \rho_\beta(E_{\alpha \mu}) \ .
\ea
We have replaced the energy denominator by a delta function (and
thereby neglected the principal-value
contribution). Expression~(\ref{r11d}) links only classes at the same
energy. That holds for all terms generated in the perturbation
expansion. As a consequence, the resulting transport equation relates
only the mean occupation probabilities $P_{E \alpha}(t)$ of states in
class $\alpha$ at energy $E$ with each other, with $E = E_{\alpha_0
  \mu_0}$ defined by the initial condition~(\ref{r9}).

Resumming the Taylor expansion of $c_{\alpha \mu} (t_1)$ gives
for expression~(\ref{r11b})
\ba
- \frac{1}{2 \hbar} \Gamma^\downarrow_{E \alpha} P_{E \alpha}(t) \ .
\label{r11e}
\ea
Here $\Gamma^\downarrow_{E \alpha}$ is the ``spreading width'' for
states in class $\alpha$ at energy $E$ defined by
\ba
\Gamma^\downarrow_{E \alpha} = 2 \pi \sum_\beta V^2_{\alpha \beta} \rho_\beta(E)
\ ,
\label{r12}
\ea
with $\rho_\beta(E)$ the average level density of states in class
$\beta$ at energy $E$. Combining the result~(\ref{r11e}) with the
corresponding contribution from the last term in Eq.~(\ref{r11}) gives
$- (\Gamma^\downarrow_{E \alpha} / \hbar) P_{E \alpha}(t)$. That is
the ``loss term'' which describes loss of occupation probability in
class $\alpha$ due to scattering into classes $\beta \neq \alpha$.

To show that terms involving $\sum_\nu t^{- l} \Delta^{- l}_\nu$ with
$l \geq 1$ in Eq.~(\ref{r11c}) are negligible, we compare the terms
with $l = 0$ and $l = 1$. According to Eq.~(\ref{r11d}) the term with
$l = 0$ is given by $\pi \hbar \rho_\beta$. For the term with $l = 1$
we have
\ba
\frac{1}{t} \bigg\langle \sum_\nu \frac{1}{\Delta^2_\nu} \bigg\rangle
&=& \frac{1}{t} \bigg\langle \sum_\nu \frac{i^2 \hbar^2}{(E_{\alpha \mu}
  - E_{\beta \nu})^2} \bigg\rangle \nonumber \\
&& \approx - i \frac{\pi \hbar^2}{t} \bigg\langle \frac{{\rm d}
  \rho_\beta(E)} {{\rm d} E} \bigg|_{E_{\alpha \mu}} \bigg\rangle\ . 
\label{r12a}
\ea
The angular bracket in the last line denotes the average over
$E_{\alpha \mu}$. Criterion~(\ref{r10a}) implies that the
term~(\ref{r12a}) is negligible compared to the term with $l =
0$. Terms of higher order $l$ yield higher-order derivatives of
$\rho_\beta$ and are negligible, too.

Contraction of the factor $\tilde{V}(t)$ with the factor immediately
to its left gives for the first term on the right-hand side of
Eq.~(\ref{r11})
\ba
\label{r12b}
&& \frac{1}{\hbar^2} \sum_\beta V^2_{\alpha \beta} \sum_{\mu \nu} \int_0^t
      {\rm d} t_1 \ \exp \{ i (E_{\alpha \mu} - E_{\beta \nu}) (t - t_1) \}
      \nonumber \\
      && \qquad \times \bigg\langle c^*_{\beta \nu}(t_1) c_{\beta \nu}(t)
      \bigg\rangle \ .
\ea
Proceeding as in Eq.~(\ref{r11c}) we obtain
\ba
\label{r12c}
\frac{1}{2} \frac{2 \pi}{\hbar} \rho_\alpha(E) \sum_\beta V^2_{\alpha
  \beta} P_{E \beta}(t) \ .
\ea
Combined with the corresponding contribution from the last term in
Eq.~(\ref{r11}) that gives $(2 \pi / \hbar) \rho_\alpha \sum_\beta
V^2_{\alpha \beta} P_{E \beta}(t)$. That is the ``gain term'' which
describes gain of occupation probability in class $\alpha$ due to
scattering out of classes $\beta \neq \alpha$.

The sum of gain and loss terms gives the Markovian transport equation
\ba
\dot{P}_{E \alpha} &=& \sum_\beta \frac{2 \pi}{\hbar} \rho_\alpha(E)
V^2_{\alpha \beta} P_{E \beta} \nonumber \\
&& \qquad - P_{E \alpha} \sum_\beta \frac{2 \pi}{\hbar}
V^2_{\beta \alpha} \rho_\beta(E) \ .
\label{r13}
\ea
We have used $V^2_{\alpha \beta} = V^2_{\beta
  \alpha}$. Eq.~(\ref{r13}) describes the time evolution of the mean
occupation probabilities $P_{E \alpha}(t)$ of states in class $\alpha$
at excitation energy $E = E_{\alpha \mu}$ defined by the initial
condition~(\ref{r9}). It is easily checked that total probability is
conserved, i.e., that Eq.~(\ref{r5}) holds. Equilibrium, characterized
by $\dot{P}_\alpha = 0$ for all $\alpha$, is attained when $P_\alpha =
C \rho_\alpha$ for all classes $\alpha$, with $C$ a normalization
constant independent of $\alpha$. On physical grounds we expect the
system to attain equilibrium for large times. Whether and how quickly
that actually happens depends upon the couplings between classes of
states and is not investigated here.

\subsection{Discussion}
\label{dis}

Eq.~(\ref{r13}) is Markovian, i.e., the time evolution of the
occupation probabilities $P_{E \alpha}(t)$ is independent of the
previous history of the system and depends only on the values of the
occupation probabilities $P_{E \alpha(t)}$ at time $t$. That is a
consequence of criterion~(\ref{r10a}). Moreover, the transport
equation~(\ref{r13}) is not only Markovian but also completely
on-shell. That follows from Eq.~(\ref{r11d}).

In the probability transport equation~(\ref{r13}), the gain (the loss)
of occupation probability in class $\alpha$ due to transport into (out
of) that class from (into) other classes at the same energy is
governed by rates (changes of occupation probability per unit
time). These rates are
\ba
R_{\alpha \rightarrow \beta} &=& (2 \pi / \hbar) V^2_{\alpha \beta} \rho_\beta
\ , \nonumber \\
R_{\beta \rightarrow \alpha} &=& (2 \pi / \hbar) V^2_{\beta \alpha}
\rho_\alpha \ .
\label{r14}
\ea
Here $R_{\alpha \rightarrow \beta}$ is the rate for decay of the
states in group $\alpha$ into states in group $\beta$, and conversely
for $R_{\beta \rightarrow \alpha}$. The two rates are related by
detailed balance, $\rho_\alpha R_{\alpha \rightarrow \beta} =
\rho_\beta R_{\beta \rightarrow \alpha}$. The expressions~(\ref{r14})
for the rates have the form of Fermi's Golden Rule, suggesting that
they can be calculated in lowest order of perturbation theory (even
though the transport equation~(\ref{r13}) has been derived to all
orders). We now show that this is indeed the case.

In lowest order of time-dependent perturbation theory, the average
time-dependent probability $P_{\alpha \mu; \beta}(t)$ for decay of a
given state $(\alpha, \mu)$ into any one of the states $(\beta, \nu)$
is given by

\ba
P_{\alpha \mu; \beta}(t) &=& \frac{1}{\hbar^2} \int_0^t {\rm d} t_1 \int_0^t
{\rm d} t_2 \nonumber \\
&& \times \sum_\nu \bigg\langle \tilde{V}_{\beta \nu; \alpha \mu}(t_1)
\tilde{V}^*_{\beta \nu; \alpha \mu}(t_2) \bigg\rangle \nonumber \\
&=& V^2_{\beta \alpha} \sum_\nu 4 \bigg\langle \frac{\sin^2[(E_{\beta \nu}
    - E_{\alpha \mu}) t / (2 \hbar)]}{(E_{\beta \nu} - E_{\alpha \mu})^2}
\bigg\rangle \nonumber \\
&=& V^2_{\beta \alpha} \int {\rm d} E_{\beta \nu} \rho_\beta \nonumber \\
&& \qquad \times 4 \bigg\langle \frac{\sin^2[(E_{\beta \nu}
    - E_{\alpha \mu}) t / (2 \hbar)]}{(E_{\beta \nu} - E_{\alpha \mu})^2}
\bigg\rangle \ .
\label{r15}
\ea
In the last line we have used the continuum approximation and have
replaced $\sum_\nu \to \int {\rm d} E_{\beta \nu} \rho_\beta$.

Two scenarios exist for evaluating the last line of Eq.~(\ref{r15}).
(i) For sufficiently large values of $t$, the integrand differs
essentially from zero only in a narrow energy interval centered at
$E_{\alpha \mu}$. In that interval, $\rho_\beta$ is approximately
constant. The remaining integral can be done and yields $2 \pi t /
\hbar$. Thus,
\ba
\lim_{t \to \infty} \frac{1}{t} P_{\alpha \mu; \beta}(t) = \frac{2 \pi}{\hbar}
V^2_{\alpha \beta} \rho_\beta \ .
\label{r16}
\ea
The right-hand side is equal to $R_{\alpha \to \beta}$ as given in
Eq.~(\ref{r14}).  However, expression~(\ref{r16}) holds only for $t
\to \infty$.  (ii) For values of $t$ relevant for the transport
equation~(\ref{r13}), condition~(\ref{r10a}) states that the level
density $\rho_\beta$ can be considered to be independent of energy and
can, thus, be taken out from under the integration. That gives
\ba
\frac{1}{t} P_{\alpha \mu; \beta}(t) = \frac{2 \pi}{\hbar} V^2_{\alpha \beta}
\rho_\beta \ ,
\label{r16a}
\ea
in full accord with Eq.~(\ref{r14}). Thus, while the standard
calculation of the rate (Eq.~(\ref{r16})) applies for very large times
only, expression~(\ref{r16a}) holds for values of $t$ that are
relevant for the transport equation. Needless to say,
expression~(\ref{r16a}) does not hold for very small times (where it
is not actually used either). Put differently, expression~(\ref{r16a})
would apply for all values of $t$ only if $\rho_\beta$ were simply
constant. That is impossible. The actual spectrum of the states $|
\beta \nu \rangle$ has a finite range $\Delta
E$. Condition~(\ref{r10a}) amounts to the requirement that $\Delta E
\gg \hbar / t$ for the times characteristic of the transport equation,
and that $\rho_\beta$ be very smooth within the energy interval
$\Delta E$.

The characteristic times of the transport equation~(\ref{r13}) are
given by the rates~(\ref{r14}). Condition~(\ref{r10a}) can, therefore,
be formulated simply by saying that for all classes $\alpha$, the
spreading width $\Gamma^\downarrow_\alpha$ must be small compared to
the energy interval $E_\alpha$ over which the average level density
$\rho_\alpha$ changes significantly.

In summary, the combination of our statistical assumptions~(\ref{r1})
and of the conditions~(\ref{r10a}) implies the Markovian, on-shell
transport equation~(\ref{r13}). It is legitimate to calculate the
rates $R_{\alpha \to \beta}$ in that equation perturbatively and for
$t \to \infty$, even though the rate equation~(\ref{r13}) is not
perturbative, and the rates are actually used for finite times.

Obviously the transport equation~(\ref{r13}) holds separately for
classes of states that carry different conserved quantum numbers.

\section{Transport Equation for Driven Systems}
\label{time}

In addition to the residual interaction $V$ we now account also for
the dipole interaction $W(t)$ in Eq.~(\ref{r0}). The time dependence
of $W(t)$ has two sources. First, $W(t)$ sets in at time $t = 0$ and
terminates at time $T$. Second, $W(t)$ oscillates harmonically with
frequency $\omega_0$. We illustrate these statements for the examples
in Section~\ref{exam}.

For the laser-nucleus interaction, $T$ is the duration time of the
laser pulse, and $\hbar \omega_0$ equals the mean photon energy in the
laser pulse. We confine ourselves to the essentials and suppress all
details due to the interaction with the electromagnetic field. These
are given in Ref.~\cite{Pal20}. We also suppress the (small)
fluctuations of the actual photon frequencies in the laser pulse
around the mean value $\omega_0$. In the interaction representation,
the time dependence of the dipole matrix element is given by Eq.(16)
of Ref.~\cite{Pal20},
\ba
\label{r17}
\theta(t) \Theta(T - t) \langle \alpha \mu | W | \beta \nu \rangle
\exp \{ i ( E_{\alpha \mu} \pm \hbar \omega_0 - E_{\beta \nu}) t / \hbar
\} \ . \nonumber \\
\ea
Here $\Theta(t)$ and $\Theta(T - t)$ are Heaviside functions
describing beginning and end in time of the laser-nucleus interaction,
and $W$ is independent of time. The term $\pm \hbar \omega_0$ in the
exponent (absent in Eq.~(\ref{r7})) corresponds, respectively, to
dipole absorption (plus sign) or to induced dipole emission (minus
sign) in the transition $(\alpha \mu) \to (\beta \nu)$.

For atoms, we assume that the two-dimensional trap has the shape of a
harmonic oscillator, with confining potential proportional to $\sum_k
(x^2_k + y^2_k)$. Here $(x_k, y_k)$ with $k = 1, \ldots, N$ are the
Cartesian coordinates of the $N$ atoms.  During the time $T$ the
confining potential is subject to externally imposed dipole
oscillations, $\sum_k (x^2_k + y^2_k) \to \sum_k [(x^2_k + y^2_k) +
  \xi (x_k - y_k) \sin (\omega_0 t)]$ with strength $\xi$. In the
interaction representation, the time dependence of the dipole matrix
element $W(t)$ also has the form of Eq.~(\ref{r17}).

In both cases, the presence of an external driving force manifests
itself in the time dependence of the matrix elements~(\ref{r17}) which
differs from that of the matrix elements of $V$ in Eq.~(\ref{r7}). The
increment $\hbar \omega_0$ is positive or negative depending on
whether energy is added to or taken from the system in the transition
$\alpha \to \beta$, and conversely for $\beta \to \alpha$. The
time-independent part of the operator $W$ with matrix elements
$\langle \alpha \mu | W | \beta \nu \rangle$ represents the
parity-changing dipole interaction, a sum of single-particle
operators. The arguments of Section~\ref{mar} show that the
non-vanishing matrix elements of $W$ are zero-centered Gaussian random
variables with second moments given by
\ba
&& \bigg\langle \langle \beta \nu | W | \alpha \mu \rangle \langle
\beta' \nu' | W | \alpha' \mu' \rangle \bigg\rangle = \bigg(
\delta_{\alpha \alpha'} \delta_{\mu \mu'} \delta_{\beta \beta'} \delta_{\nu \nu'}
\nonumber \\
&& \qquad \qquad \qquad + \delta_{\alpha \beta'} \delta_{\mu \nu'}
\delta_{\beta \alpha'} \delta_{\nu \mu'} \bigg) W^2_{\beta \alpha} \ .
\label{r17a}
\ea
Again, we have $W^2_{\beta \alpha} = W^2_{\alpha \beta}$.

In deriving the transport equation we follow the steps taken in
Sections~\ref{mark}, \ref{rate}, and in the Appendix. The elements of
$V$ and of $W(t)$ are uncorrelated zero-centered Gaussian random
variables. The contraction rules apply separately for the elements of
$V$ and of $W$. The time dependence of $W(t)$ causes a
change of Eq.~(\ref{r11d}) which now reads
\ba
\label{r17b}
\bigg\langle \sum_\nu \frac{i \hbar}{E_{\alpha \mu} \pm \hbar \omega_0
  - E_{\beta \nu}} \bigg\rangle \approx \pi \hbar \rho_\beta(E_{\alpha \mu}
\pm \hbar \omega_0) \ .
\ea
That shows that the second moment $W^2_{\alpha \beta}$ connects
classes of states that differ in energy by $\pm \hbar \omega_0$. In
averaging over the Gaussian-distributed matrix elements of $W$, we
again use conditions~(\ref{r10a}). The resulting contraction rules are
the same as for the matrix elements of $V$.

To derive the transport equations, we start from Eq.~(\ref{r11}). The
perturbation series for $\dot{c}_\alpha(t)$ has the form of
Eq.~(\ref{r11a}), with the replacement $\tilde{V}(t) \to \tilde{V}(t)
+ W(t)$ in every term of the series, and correspondingly for
$c^*$. The first factor in the expansion of $\dot{c}_ {\alpha \mu}(t)$
may either be a matrix element of $V$, or a matrix element of $W$. In
the first case, conditions~(\ref{r10a}) imply that a non-vanishing
result is obtained only if either the second factor in $\dot{c}$ or
the first factor in $c^*$ is also a matrix element of $V$. Proceeding
as in Section~\ref{deri} we arrive at the right-hand side of
Eq.~(\ref{r13}). In the second case, a nonvanishing result is obtained
only if either the second factor in $\dot{c}$ or the first factor in
$c^*$ is also a matrix element of $W$. Proceeding as in
Section~\ref{deri} but now with regard to the matrix elements of $W$,
and taking account of Eq.~(\ref{r17b}), we find that the resulting
contribution to the transport equation connects classes of states that
differ in energy by $\pm \hbar \omega_0$. As in Eq.~(\ref{r13}) we
account in the resulting transport equation explicitly for the energy
$E$, denoting the mean occupation probability at energy $E$ of the
states in class $\alpha$ at time $t$ by $P_{E, \alpha}(t)$. The
transport equation is
\ba
\label{r17c}
&& \dot{P}_{E \alpha} = \sum_\beta \frac{2 \pi}{\hbar} \rho_{E \alpha}
V^2_{E \alpha, E \beta} P_{E \beta} - P_{E \alpha} \sum_\beta
\frac{2 \pi}{\hbar} V^2_{E \beta, E \alpha} \rho_{E \beta} \nonumber \\
&& \qquad + \theta(t) \Theta(T - t) \bigg[ \sum_{\beta} \sum_{j = \pm 1}
  \frac{2 \pi}{\hbar} \rho_{E \alpha} W^2_{E \alpha, E + j \hbar \omega_0 \beta}
  \nonumber \\
  && \qquad \qquad \times P_{(E + j\hbar \omega_0) \beta} \nonumber \\
&& \qquad - P_{E \alpha} \sum_{\beta} \sum_{j = \pm 1} \frac{2 \pi}{\hbar}
     \rho_{(E + j \hbar \omega_0) \beta} W^2_{E \alpha, (E \pm j \hbar
       \omega_0) \beta} \bigg] \ .
\ea
The terms in the first line correspond to Eq.~(\ref{r13}). The terms
in the second and third line feed the occupation probability in class
$(E, \alpha)$ either because of dipole excitation of states in class
$(E - \hbar \omega_0, \beta)$ or because of induced dipole emission of
states in class $(E + \hbar \omega_0, \beta)$, and conversely for the
loss terms in the fourth line. Again, the rates are related by detailed
balance, and Eq.~(\ref{r17c}) conserves occupation probability.

The rates for internal equilibration and for driven excitation must be
determined individually for each system. Two examples show how to
proceed. In Ref.~\cite{Kob20} the imaginary part of the optical model
for elastic scattering of nucleons by nuclei was used to determine the
strength of the two-body interaction. In combination with level
densities obtained from the nuclear shell model, that yielded the
rates for internal equilibration. Similarly, in Ref.~\cite{Pal20} the
rate for laser-induced dipole excitation was calculated for a laser
pulse carrying $N$ photons with mean energy $\hbar \omega_0$ and
duration time $\tau$ using nuclear level densities and the dipole
operator. It is important to realize that quite generally, rates may
be calculated perturbatively using Fermi's Golden Rule even though
Eq.~(\ref{r17c}) itself is non-perturbative. That was demonstrated in
Section~\ref{dis} and in Ref.~~\cite{Pal20}.

In fermionic many-body systems, the average level density rises
steeply with excitation energy~\cite{Bet37}. The steep rise, valid for
an infinitely deep single-particle potential, is modified when the
finite depth of the potential is taken into account. For the nuclear
case that is demonstrated in Refs.~\cite{Pal13, Pal13a}. With
increasing energy, the level density rises up to a maximum value $E_0$
of the energy whereupon it decreases. As long
as the level density increases, detailed balance implies that dipole
absorption is stronger than stimulated dipole emission, and the
transport equation~(\ref{r17c}) drives the system to higher excitation
energy.  At $E = E_0$, however, the rates for induced dipole
absorption and for stimulated dipole emission are equal while for $E >
E_0$ stimulated emission prevails. Thus, $E_0$ marks the maximum
energy the system can absorb. The occupation probability of the states
is driven towards a saturation value at energy $E_0$ where it becomes
stationary. That is illustrated in Refs.~\cite{Pal15, Kob20} where
variants of Eqs.~(\ref{r17c}) have been applied to the laser-nucleus
interaction. Thermodynamically, $E_0$ corresponds to infinite
temperature.

Actually, dipole absorption is limited not by thermodynamic saturation
but by evaporation. The system spills particles (nuclei or atoms) and,
thereby, loses energy. The process is caused by the finite depth of
the single-particle potential. Evaporation sets in at energies much
lower than $E_0$. For nuclei, evaporation of $s$-wave neutrons
dominates. It is described by the Weisskopf estimate~\cite{Wei37}.
The estimate is based on statistical arguments involving average level
densities. To account for evaporation in the present framework,
Eqs.~(\ref{r17c}) are, in effect, replaced by a chain of coupled
equations, each describing the system after the loss of $j = 0, 1, 2,
\ldots$ particles. Each such equation has the form of
Eqs.~(\ref{r17c}) but carries additional terms involving the Weisskopf
estimate that account for the feeding (loss) of probability due to
particle evaporation from the previous (from the present) member of
the chain. Such equations can also be derived in the present framework
but that is not shown here. Equations of that type have been used, for
instance, in Refs.~\cite{Pal15, Kob20}. For atoms the trap potential
has finite depth as well, and evaporation is expected to work
similarly. In nuclei, dipole excitation may also be limited by
fission.

In Eqs.~(\ref{r17c}) we have not displayed explicitly quantum numbers
like parity, spin, or total angular momentum. The dipole interaction
changes parity, and it changes spin or angular momentum by one
unit. Thus, Eqs.~(\ref{r17c}) actually couple all classes with these
quantum numbers. The number of coupled equations is large, especially
when particle evaporation is also taken into account. For that reason,
it may be advisable to use transport equations for occupation
probabilities averaged over some or all these quantum
numbers as done in Refs.~\cite{Pal15, Kob20}.

\section{Nuclear Dipole Absorption}
\label{nda}

In applications, the formalism developed in previous sections may
require some modification. We demonstrate that for the case of
laser-induced nuclear dipole absorption. We consider two cases that
differ in the intensity of the laser pulse or, equivalently, in the
rate of dipole excitation. If that rate is small in comparison with
the rate for on-shell relaxation, each photon absorption process is
followed by complete internal relaxation. That defines the
``quasiadiabatic'' regime investigated in Refs.~\cite{Pal14, Pal15}.
The rate for dipole absorption used in these papers was later
shown~\cite{Pal20} to follow from the the Brink-Axel
hypothesis~\cite{Bri55, Axe62}. If, on the other hand, the rates for
dipole absorption and for internal relaxation are approximately equal,
both processes occur simultameously. That defines the ``sudden''
regime studied in Ref.~\cite{Kob20}. Then the Brink-Axel hypothesis
must be modified. We show how these alternatives fit into our general
framework.

\subsection{Quasiadiabatic Regime}
\label{adia}

Here the rate for photon absorption is small compared to the rate for
equilibration. After every absorption of a photon, the nucleus
equilibrates. It is not necessary to consider classes of states
carrying different particle-hole numbers. The class label $\alpha$
stands for spin $J$ and parity $\pi$. The states in class $(J, \pi)$
are eigenstates $| \mu \rangle_{J \pi}$ of the nuclear Hamiltonian
$H_N$. The level density is $\rho_{J \pi}(E)$. In the transport
equations~(\ref{r17c}), the terms on the right-hand side of the first
line are absent. The role of the operator $W(t)$ in Eq.~(\ref{r0}) is
taken by the dipole approximation to the full photon-nucleus
interaction Hamiltonian ${\cal H}(t)$, the product of the nuclear
current operator and the quantized radiation field.

For a laser pulse carrying $N$ photons with mean energy $\hbar
\omega_0$, the rate $R$ for photon absorption from an arbitrary initial
nuclear state $| i J_i \rangle$ with energy $E_i$ and spin $J_i$ has
been calculated perturbatively in Ref.~\cite{Pal20} with the help of
the Brink-Axel hypothesis. As stated in Section~\ref{las}, the
hypothesis says that dipole absorption by the state $| i J_i \rangle$
populates the normalized dipole modes $| d(i) J_f \rangle$ pertaining
to that state. These carry spins $J_f = J_i, J_i \pm 1$ and, assuming
degeneracy, have energy $E_d = \langle d(i) | H_N | d(i) \rangle$. The
rate $R$ is~\cite{Pal20}
\ba
R &=& N \frac{c}{9 \pi} \frac{e^2}{\hbar c} \frac{\Gamma^\downarrow \sigma}
{(E_i + \hbar \omega_0 - E_d)^2 + (1 / 4) (\Gamma^\downarrow)^2} \nonumber \\
&& \times 4 \pi \alpha \frac{(\hbar \omega_0)^3}{(\hbar c)^3} \sum_{J_f} |
\langle i J_i || r Y_1 || d(i) J_f \rangle |^2 \ .
\label{r20}
\ea
Here $\alpha \ll 1$ is the aperture of the laser pulse, $\sigma
\approx 10$ keV is the spread in energy of the laser pulse around its
mean value $\hbar \omega_0$, the sum over $J_f$ extends over all spin
values that can be reached via dipole absorption from the initial
state $| i \rangle$ with spin $J_i$, and $| \langle i J_i || r Y_1 ||
d(i) J_f \rangle |^2$ is the square of the reduced dipole matrix
element. The operator $r Y_1$ comprises a summation over neutrons and
protons with their effective charges, written in units of $e$. The
Lorentzian in Eq.~(\ref{r20}) stems from that fact that the dipole
mode is spread over the eigenstates $| \mu \rangle$ of $H_N$ with a
Lorentzian distribution centered at $E_d$. That follows from averaging
over a random-matrix model for $H_N$, see, for instance,
Ref.~\cite{Wei11}. The spreading width is
\ba
\Gamma^\downarrow = 2 \pi \bigg\langle | \langle \mu | H_N | i(d) \rangle
|^2 \bigg\rangle \rho(E_d) \ .
\label{r19}
\ea
Here $\rho(E_d)$ is the nuclear level density at energy $E_d$. We have
suppressed the dependence on initial and final spins. That is in line
with the use made of Eqs.~(\ref{r20}) and (\ref{r19}) in
Refs.~\cite{Pal20}. There the square of the reduced dipole matrix
element in Eq.~(\ref{r20}) is estimated via the dipole sum rule. Then
the rate $R$ in Eq.~(\ref{r20}) becomes independent of the initial
state $| i \rangle$ and depends only on the difference $E_d - E_i$. It
is assumed that within some domain of excitation energies and nuclear
mass numbers, that difference is approximately constant, i.e.,
independent of the initial nuclear state $| i \rangle$. Likewise it is
assumed that the spreading width $\Gamma^\downarrow$ is independent of
excitation energy and mass number. These assumptions are in line with
the asssumed universality of the Brink-Axel hypothesis. As a result
the rate $R$ for dipole excitation from any initial nuclear state is
universal. The rates for stimulated absorption follow from detailed
balance. In Refs.~\cite{Pal14, Pal15} the transport
equations~(\ref{r17c}) have been simplified further by suppressing the
quantum numbers $(J, \pi)$.  That is justified by the Gaussian
dependence of $\rho_{J \pi}(E)$ on $J$ and because $\rho_{J \pi}(E)$
is nearly independent of $\pi$. Combined with plausible estimates for
$R$ (that have later been confirmed in Ref.~\cite{Pal20}), the
resulting rate equations have been used in Refs.~\cite{Pal14, Pal15}
to calculate laser-induced multi-photon absorption in nuclei in the
quasiadiabatic regime.

The dipole strength is spread over an energy interval of Lorentzian
form with width $\Gamma^\downarrow$. In a time-dependent picture,
spreading happens during a time of order $\hbar / \Gamma^\downarrow$.
We refer to that process as to equilibration because it happens within
a single class of states defined by spin and parity. In the
quasiadiabatic regime, the time scale $R^{- 1}$ for dipole absorption
must be larger than that time, $R^{- 1} \geq 2 \pi \hbar /
\Gamma^\downarrow$. Introducing the dipole width $\Gamma_{\rm dip} =
\hbar R$ we write that condition in intuitively appealing form as
$\Gamma^\downarrow \geq 2 \pi \Gamma_{\rm dip}$. That is consistent
with both, the general discussion of the derivation and applicability
of rates in Section~\ref{dis} and with a similar, more restricted
derivation of expression~(\ref{r20}) for $R$ in Ref.~\cite{Pal20}. In
both cases it is shown that Eq.~(\ref{r20}) actually applies for
times $t \geq 2 \pi \hbar / \Gamma^\downarrow$.

To see that the contraction rules and the ensuing Markov approximation
also apply in the present case we note that the role played by the
level density in the rate expressions~(\ref{r14}) is, in the case of
Eq.~(\ref{r20}), taken by the Lorentzian. Condition~(\ref{r10a}) for
the validity of the Markov approximation, summarily written as $\hbar
/ (t \Delta E) \ll 1$ where $\Delta E$ is the energy interval over
which any level density changes significantly, changes to $\hbar / (t
\Gamma^\downarrow) \ll 1$. Replacing $\hbar / t \to \Gamma_{\rm dip}$
we obtain $2 \pi \Gamma_{\rm dip} < \Gamma^\downarrow$. That agrees
with the condition derived in the previous paragraph. It implies that
the contraction rules do apply and that the Markov form of the
transport equation approximately holds in the quasiadiabatic
regime. The conditions for validity of the Markov approximation and of
the adiabatic regime are seen to coincide. That is true even though
$\Gamma^\downarrow$ is, as a rule, significantly smaller than $\Delta
E$, and the condition $\hbar / (t \Gamma^\downarrow) \ll 1$ is more
stringent.

The definition $\Gamma^\downarrow \geq 2 \pi \Gamma_{\rm dip}$ of the
adiabatic regime must obviously be understood as an asymptotic
condition. The more closely $\Gamma_{\rm dip}$ approaches
$\Gamma^\downarrow$ from below, the bigger are the expected deviations
from the Markov approximation. The ensuing limitation of the
quasiadiabatic approach is circumvented by going to the sudden regime
in Section~\ref{sudd} where the Markov approximation is reestablished
albeit on a different time scale. The calculations in
Refs.~\cite{Pal14, Pal15} have used the transport equation for the
quasiadiabatic regime but have been done, for instance, for
$\Gamma^\downarrow = 5$ MeV, $\Gamma_{\rm dip} = 5$ MeV, violating the
condition $\Gamma^\downarrow \geq 2 \pi \Gamma_{\rm dip}$. The
comparison with results~\cite{Kob20} for the sudden regime shows that
deviations from the Markov approximation set in rather slowly.

The Brink-Axel hypothesis is well established for excitation energies
in the MeV range~\cite{Mar17}. It is not clear, however, to what
extent it actually holds for very highly excited initial states. It is
conceivable that with increasing excitation energy, the hypothesis
gradually loses validity. Then the spread $\Delta E$ of the dipole
strength would increase. At large excitation energies the GDR is
primarily observed experimentally via gamma decay of compound nuclei
formed in heavy-ion collisions. It is found that the GDR is
substantially broadened or disappears altogether~\cite{Yos90, San20,
  Bor91}. That is ascribed either to the large angular-momentum values
involved in a heavy-ion induced reaction, to neutron evaporation, or
to a significant intrinsic dynamical broadening of the GDR. The first
cause is irrelevant here because laser-induced nuclear dipole
excitation does not populate states with large spin values. Neutron
evaporation is accounted for explicitly in the coupled transport
equations of Ref.~\cite{Pal15, Kob20}. We, therefore, focus attention
on the possibility that the GDR is significantly broadened
dynamically.

To account for that case we calculate the dipole absorption rate under
a weaker and more general assumption than used in the Brink-Axel
hypothesis. As in Section~\ref{time} we assume that the reduced
nuclear dipole matrix elements connecting any initial state $| i J_i
\rangle$ with any final state $| f J_f \rangle$ are zero-centered
Gaussian-distributed random variables, and that the dipole strength of
any initial state $| i J_i \rangle$ is on average distributed
uniformly over the final states $| f J_f \rangle$ in an energy
interval $\Delta E$ that is considerably larger than the spreading
width $\Gamma^\downarrow$ characteristic of the ground-state
regime. Then equilibration, governed by the time scale $\hbar / \Delta
E$, is much more rapid than in the case of the Lorentzian in
Eq.~(\ref{r20}), and the conditions for the validity of the Markov
approximation are fulfilled more readily. It is straightforward to
repeat the calculation in Section~VI of Ref.~\cite{Pal20} for that
case. The result for the rate is
\ba
\tilde{R} &=& N \frac{c}{9 \pi} \frac{e^2}{\hbar c} 4 \pi \alpha 
\frac{(\hbar \omega)^3}{(\hbar c)^3} \sigma \nonumber \\
&& \times 2 \pi \rho(E_i + \hbar \omega) \sum_{J_f} | \langle i J_i ||
r Y_1 || f J_f \rangle |^2 \ .
\label{r24}
\ea
Here $\rho(E_i + \hbar \omega)$ is the density of states at energy
$(E_i + \hbar \omega)$, assumed to be the same for all three spin
values $J_f$. The last line in Eq.~(\ref{r24}) has the standard form
of a rate.

We compare the rates~(\ref{r20}) and (\ref{r24}). The sum over $J_f$
in expression~(\ref{r20}) exhausts the dipole sum rule. To estimate
the reduced matrix elements in Eq.~(\ref{r24}), we observe that the
sum $\sum_{f J_f} | \langle i J_i || r Y_1 || f J_f \rangle |^2$
exhausts the dipole sum rule if it extends over states $f$ within the
energy interval $\Delta E$. Thus $\rho(E_i + \hbar \omega) \sum_{J_f}
| \langle i J_i || r Y_1 || f J_f \rangle |^2$ is approximately equal
to $\sum_{J_f} | \langle i J_i || r Y_1 || d(i) J_f \rangle |^2 /
\Delta E$, and we obtain
\ba
\frac{\tilde{R}}{R} \approx \frac{2 \pi \Gamma^\downarrow}{\Delta E} \ . 
\label{r25}
\ea
Significant deviations from the Brink-Axel hypothesis occur if that
ratio is of order $10^{- 1}$ or less. The photon absorption
probability is reduced by that factor. A possible experimental signal
for that to happen is a deceleration of the laser-induced photon
absorption process with increasing nuclear excitation energy.

\subsection{Sudden Regime}
\label{sudd}

In the sudden regime the dipole rate is at least as large as or even
bigger than the rate for nuclear equilibration. The statistical
approach of Section~\ref{stat} is based on the assumption that there
exists a set of quickly equilibrating subsystems. The time scale for
internal equilibration must be short in comparison with both, the
dipole rate and the rate for nuclear relaxation.

To identify these subsystems we recall that in the ground-state
domain, the dynamical description of the GDR involves a two-step
process. First, dipole absorption populates a set of particle-hole
states. These are not completely degenerate and, thus, contribute to
the spreading of the GDR. That one-body effect is referred to as
Landau damping~\cite{Spe91}. The particle-hole states are not
eigenstates of the nuclear Hamiltonian and mix with $m$p-$m$h states
with $m > 1$. That mixing amounts to relaxation and gives rise to the
total spreading of the GDR.

In line with that picture, the subsystems in Ref.~\cite{Kob20} are
taken as classes of particle-hole states at fixed total energy. It is
assumed that within each subsystem, equilibration is much more rapid
than the mixing of different subsystems leading to relaxation. Such an
assumption is similarly used in the theory of precompound
reactions~\cite{Bla75}. Here classes of states populated in the
reaction are also classified according to the number of particles and
holes. For instance, the collision of an incident proton with a
nucleon in the target nucleus creates a two-particle one-hole state,
the next collision leads to a three-particle two-hole state, etc. It
is assumed that after each collision the $(m + 1)$p-$m$h states
equilibrate rapidly, so that level densities may be used for
describing the reaction. As a result, the sequence of collisions is
described in terms of rates, each rate connecting a class of $(m +
1)$p-$m$-hole states with the next one. The success of that
model~\cite{Bla75} supports the assumption of rapid equilibration. A
similar physical picture underlies the approach developed in
Ref.~\cite{Aga75}.

Implementation of these ideas leads straightforwardly to the transport
equations~(\ref{r17c}). The class label $\alpha$ is identified with
spin, parity, and particle-hole number combined. Equilibration within
each class is instantaneous. The remaining part of the two-body
residual nuclear interaction connects only classes with identical
quantum numbers, and it changes particle-hole number only by one
unit. That yields the terms in the first line of Eq.~(\ref{r17c}).
Emission and absorption of photons is described by the dipole
approximation to the full photon-nucleus interaction Hamiltonian
${\cal H}(t)$ mentioned above. Dipole absorption either leaves
particle-hole number unchanged or increases it by unity. Calculation
of the rates for dipole absorption yields expressions like in
Eq.~(\ref{r24}) but specified to fixed particle-hole number, see
Ref.~\cite{Kob20}. That yields the remaining two lines in
Eq.~(\ref{r17c}). The rates are normalized to the dipole sum rule.

\section{Summary and Discussion}
\label{sum}

Transport equations for autonomous driven fermionic quantum systems
are derived with the help of statistical assumptions, and of the
Markov approximation. The statistical assumptions hold if the system
consists of subsystems within which equilibration is sufficiently
fast. The Markov approximation holds if the level density in each
subsystem is sufficiently smooth in energy. From a formal point of
view, Eqs.~(\ref{r1}, \ref{r17a}) and the inequality~(\ref{r10a})
constitute sufficient conditions for the validity of Eqs.~(\ref{r13})
and (\ref{r17c}). The transport equations describe both, internal
equilibration among subsytems at equal energy and the transport of the
system to higher energy caused by the driving force. The result puts
the use of transport equations for the laser-nucleus interaction in
Refs.~\cite{Pal14, Pal15, Kob20} on a firm theoretical basis. We are
hopeful that such equations prove useful also for atoms in driven
traps.

We end with some general remarks. The result~(\ref{r13}) for internal
relaxation has the form of the Pauli master equation. That equation
goes back to the beginning of quantum mechanics. Nonetheless,
derivation and validity of Eq.~(\ref{r13}) remain a frequently
discussed and timely topic, especially for autonomous systems. The
derivation of Eq.~(\ref{r13}) usually employs coupling to a
reservoir. In that framework, relevant time scales are the decoherence
time $\tau_{\rm dec}$ and the dissipation time $\tau_{\rm diss}$, see,
for instance, Ref.~\cite{Ost17}. Here $\tau_{\rm dec}$ is the time
scale for exponential decay of the off-diagonal matrix elements of the
density matrix, and $\tau_{\rm diss}$ is the time it takes the
diagonal elements of the density matrix (i.e., our occupation
probabilities $P_{E \alpha}(t)$) to attain equilibrium. Although we
deal with an autonomous system, analogous time scales appear in our
work. In Section~\ref{rate}, nondiagonal elements of the density
matrix have been shown to disappear upon averaging over the
random-matrix ensemble. Such averaging is justified for times larger
than the internal equilibration time within each class $\alpha$ of
states. Thus, our internal equilibration time bears a close analogy to
$\tau_{\rm dec}$. Our relaxation time is identical with $\tau_{\rm
  diss}$ because it is determined by the same equation.

Our derivation of the transport equations~(\ref{r13}) and (\ref{r17a})
uses the decomposition~(\ref{r0}) of the Hamiltonian $H$ into a
shell-model part $H_0$ and the remainder. On that basis, we define
classes of states $\alpha$ and assume that within each class,
equilibration is fast. Actually, the transport equations have a wider
scope. Any system for which the states in Hilbert space can be grouped
into classes such that the states within each class interact more
strongly with each other than with the rest, so strongly, in fact,
that states within each class are thoroughly mixed prior to overall
relaxation, obeys the assumptions used in our derivation. The time
evolution of the system is then described by equations like
(\ref{r13}) and (\ref{r17a}). The label $\alpha$ refers to the classes
of strongly interacting states.
  
Our Eq.~(\ref{r13}) holds for autonomous many-body quantum systems
(systems not coupled to a reservoir). For such systems, relaxation
towards statistical equilibrium has been experimentally observed in
several areas of physics and has been interpreted with the help of a
master equation kin to Eq.~(\ref{r13}). Precompound reactions provide
an early example in nuclear physics. These show relaxation towards the
equilibrated compound nucleus, accompanied by emission of
particles~\cite{Bla75, Wei08}. A recent example for quantum
thermaliztion in atomic physics is Ref.~\cite{Kau16}. In such cases,
the derivation of a master equation without recourse to a reservoir
and, more generally, the understanding of quantum thermalization in
isolated systems, pose a problem. The derivation of the quantum
Boltzmann equation using field theory shows one solution to the
problem. Here we have shown that insights gained in random-matrix
theory also offer a solution. Arguing that equilibration within each
class $\alpha$ of states results in strongly mixed states as described
by random-matrix theory, we have obtained Eq.~(\ref{r13}) by averaging
over the resulting ensemble of random matrices. We believe that our
approach clearly identifies the physical assumptions needed for
Eq.~(\ref{r13}) and for the more general Eq.~(\ref{r17c}) to hold.

{\bf Acknowledgment.} The author is grateful to A. Palffy for a
careful reading of several versions of the paper, and for many helpful
suggestions and constructive remarks.

\section*{Appendix}

We show that contraction of the first factor $\tilde{V}_{\alpha \mu,
  \beta \nu}(t)$ in the expansion~(\ref{r11a}) of $\dot{c}_{\alpha
  \mu}$ with any other factor $\tilde{V}$ in that expansion except the
second, yields terms that are negligible by the
condition~(\ref{r10a}). With $c^*_{\alpha \mu}$ written in the manner
described below Eq.~(\ref{r11a}), the same arguments can be used to
show that contraction of the first factor $\tilde{V}_{\alpha \mu,
  \beta \nu}(t)$ in $\dot{c}_{\alpha \mu}$ with any factor $\tilde{V}$
in $c^*_{\alpha \mu}$ except the right-most one, are likewise
negligible. That is not done here, and we confine ourselves to the
expansion~(\ref{r11a}) of $\dot{c}_{\alpha \mu}$.

In carrying out the time integrations we make use of Eq.~(\ref{r11c})
where we replace the last straight bracket by unity. For every
function $F(t)$ that possesses a Taylor expansion in $t$, we then have
\ba
\label{a1}
\sum_\nu \int_0^t {\rm d} t_1 \ \exp \{ \Delta_\nu t_1) \} F(t_1)
\approx \sum_\nu \frac{1}{\Delta_\nu} \exp \{ \Delta_\nu t \} F(t) \ .
\nonumber \\
\ea
We use Eq.~(\ref{a1}) also in cases where the summation over $\nu$ is
performed somewhere later in the calculation, perhaps in conjunction
with additional factors $1 / \Delta_\nu$.

We contract the first factor $\tilde{V}(t)_{\alpha \mu, \beta \nu}$ in
the expansion~(\ref{r11a}) with the factor $\tilde{V}_{\alpha_n \mu_n,
  \beta_n \nu_n}(t_n)$ for $n \geq 2$, prior to any contraction
affecting the other factors $\tilde{V}$ in that expansion. Among the
two options of Eq.~(\ref{r8}) we first consider the case $\beta_n =
\alpha, \nu_n = \mu$. That gives
\ba
\label{a2}
&& \frac{1}{(i \hbar)^{n+1}} {\cal T} \int_0^t \prod_{k = 1}^n
      {\rm d} t_k \ \sum_{\beta \nu} V^2_{\alpha \beta}
      \bigg( \prod_{l = 1}^{n - 1} \tilde{V}(t_l) \bigg)_{\beta \nu,
        \beta \nu} \nonumber \\
      && \qquad \times \exp \{ i (E_{\alpha \mu} - E_{\beta \nu})
      (t - t_n) / \hbar \} c_{\alpha \mu}(t_n) \ .
\ea
Here we must have $n \geq 3$ because $V$ is nondiagonal in class index.
Carrying out the time integration over $t_n$ we obtain
\ba
\label{a3}
&& \frac{1}{(i \hbar)^n} {\cal T} \int_0^t \prod_{k = 1}^{n - 1}
      {\rm d} t_k \ \sum_{\beta \nu} V^2_{\alpha \beta}
      \bigg( \prod_{l = 1}^{n - 1} \tilde{V}(t_l) \bigg)_{\beta \nu,
        \beta \nu} \nonumber \\
      && \qquad \times \exp \{ i (E_{\alpha \mu} - E_{\beta \nu})
      (t - t_{n - 1}) / \hbar \} \nonumber \\
      && \qquad \times \frac{1}{E_{\alpha \mu} - E_{\beta \nu}} \
      c_{\alpha \mu}(t_{n - 1}) .
\ea
We contract $\tilde{V}(t_1)$ with $\tilde{V}(t_{n - 1})$. We show
presently that contracting either of these factors with any other
factor $\tilde{V}$ in $c$ or in $c^*$ leads to higher derivatives than
the first of the average level density $\rho_\beta$. We obtain
\ba
\label{a4}
&& \frac{1}{(i \hbar)^n} {\cal T} \int_0^t \prod_{k = 1}^{n - 1}
      {\rm d} t_k \ \sum_{\beta \gamma \nu \rho} V^2_{\alpha \beta}
      V^2_{\beta \gamma} \bigg( \prod_{l = 2}^{n - 2} \tilde{V}(t_l)
      \bigg)_{\gamma \rho, \gamma \rho} \nonumber \\ && \ \ \times \exp
      \{ i (E_{\alpha \mu} - E_{\beta \nu}) t / \hbar \} \exp \{ i
      (E_{\beta \nu} - E_{\gamma \rho}) t_1 / \hbar \} \nonumber \\
      && \ \ \times\exp \{ i (E_{\gamma \rho} - E_{\alpha \mu})
      t_{n - 1} \} \frac{1}{E_{\alpha \mu} - E_{\beta \nu}}
      \ c_{\alpha \mu}(t_{n - 1}) \ .
      \ea
We use Eq.~(\ref{r7}) and successively perform the integrations over
$t_{n - 1}$, $t_{n - 2}$, $\ldots$, $t_2$, using Eq.~(\ref{a1}). That
gives 
\ba
\label{a5}
&& \frac{1}{(i \hbar)^2} \int_0^t {\rm d} t_1 \sum_{\beta \gamma
  \nu \rho} V^2_{\alpha \beta} V^2_{\beta \gamma} \bigg( V^{n - 3}
\bigg)_{\gamma \rho, \gamma \rho} \Delta \nonumber \\
&& \ \ \times \exp \{ i (E_{\alpha \mu} - E_{\beta \nu}) t / \hbar \}
\exp \{ i (E_{\beta \nu} - E_{\alpha \mu}) t_1 / \hbar \} \nonumber \\
&& \ \ \times \frac{1}{E_{\alpha \mu} - E_{\beta \nu}} \ c_{\alpha \mu}(t_1) \ .
\ea
The factor $\Delta$ is the product of sums over inverse energy
differences. These arise in the time integrations. None of these
carries $E_{\beta \nu}$. The big round bracket is the product of the
time-independent matrices $V$ introduced in Eq.~(\ref{r1}). We
integrate over $t_1$ and use Eq.~(\ref{a1}). In Eq.~(\ref{a5}) that
yields the factor $\sum_\nu (E_{\alpha \mu} - E_{\beta \nu})^{- 2}
\propto ({\rm d} / {\rm d} E) \rho_\beta$ which is
negligible. Contraction of $\tilde{V}(t_1)$ and of $\tilde{V}(t_{n -
  1})$ in Eq.~(\ref{a2}) not with each other but with other factors
$\tilde{V}$ in $\dot{c}$ or in $c^*$ causes the occurrence of
additional terms $E_{\beta \nu}$ in exponential factors. The
associated time integrations produce higher-order derivatives of
$\rho_\beta$.

We turn to the second option in the contraction of $\tilde{V}_{\alpha
  \mu, \beta \nu}(t)$ with $\tilde{V}_{\alpha_n \mu_n, \beta_n
  \nu_n}(t_n)$ and put $\alpha_n = \alpha$, $\mu_n = \mu$. In the
expansion of $\dot{c}$ we now carry explicitly all matrix elements
$\tilde{V}(t_l)$ up to and including the one with $l = n + 1$. That
gives
\ba
\label{a6}
&& \frac{1}{(i \hbar)^{n+2}} {\cal T} \int_0^t\prod_{k = 1}^{n + 1}
      {\rm d} t_k \ \sum_{\beta \nu \gamma \rho} V^2_{\alpha \beta}
      \bigg( \prod_{l = 1}^{n - 1} \tilde{V}(t_l) \bigg)_{\beta \nu,
        \alpha \mu} \nonumber \\
      && \qquad \times \exp \{ i (E_{\alpha \mu} - E_{\beta \nu})
      (t + t_n) / \hbar \} \tilde{V}_{\beta \nu, \gamma \rho}(t_{n + 1})
      \nonumber \\
      && \qquad \times c_{\gamma \rho}(t_{n + 1}) \ .
\ea
We contract $\tilde{V}(t_1)$ with $\tilde{V}(t_{n + 1})$. Other
contraction patterns generate additional terms $E_{\beta \nu}$ in
exponentials and, thus, derivatives of higher order than the first of
$\rho_\beta$. These will not be considered. The result is
\ba
\label{a7}
&& \frac{1}{(i \hbar)^{n+2}} {\cal T} \int_0^t\prod_{k = 1}^{n + 1}
      {\rm d} t_k \ \sum_{\beta \nu \gamma \rho} V^2_{\alpha \beta}
      V^2_{\beta \gamma} \bigg( \prod_{l = 2}^{n - 1} \tilde{V}(t_l)
      \bigg)_{\gamma \rho, \alpha \mu} \nonumber \\
      && \qquad \times \exp \{ i (E_{\alpha \mu} - E_{\beta \nu})
      (t + t_n) / \hbar \} \nonumber \\
      && \qquad \times \exp \{ i (E_{\beta \nu} - E_{\gamma \rho}) (t_1
      + t_{n + 1} / \hbar \} \nonumber \\
      && \qquad \times c_{\gamma \rho}(t_{n + 1}) \ .
\ea
Integration over $t_{n + 1}$ gives
\ba
\label{a8}
&& \frac{1}{(i \hbar)^{n + 1}} {\cal T} \int_0^t\prod_{k = 1}^n
      {\rm d} t_k \ \sum_{\beta \nu \gamma \rho} V^2_{\alpha \beta}
      V^2_{\beta \gamma} \bigg( \prod_{l = 2}^{n - 1} \tilde{V}(t_l)
      \bigg)_{\gamma \rho, \alpha \mu} \nonumber \\
      && \qquad \times \exp \{ i (E_{\alpha \mu} - E_{\beta \nu}) t /
      \hbar\} \exp \{ i (E_{\beta \nu} - E_{\gamma \rho}) t_1 / \hbar \} 
      \nonumber \\
      && \qquad \times \exp \{ i (E_{\alpha \mu} - E_{\gamma \rho}) t_n
      / \hbar \} \frac{1}{E_{\beta \nu} - E_{\gamma \rho}} \nonumber \\
      && \qquad \times c_{\gamma \rho}(t_n) \ .
\ea
Proceeding as before we see that the integration over $t_1$ eventually
yields another factor $1 / (E_{\beta \nu} - E_{\gamma
  \rho})$. Combining the two denominators and summing over $\nu$ we
obtain the derivative of the level density $\rho_\beta$ which is
negligible.

\end{document}